\documentclass[12pt]{article}
\usepackage{epsf}
\usepackage{a4wide,graphicx}
\usepackage{cite}
\usepackage{subfigure}

\newcommand{\mev}{\textrm{ MeV}}
\newcommand{\be}{\begin{equation}}
\newcommand{\ee}{\end{equation}}
\newcommand{\ba}{\begin{eqnarray}}
\newcommand{\ea}{\end{eqnarray}}
\newcommand{\nn}{\nonumber}
\newcommand{\PL}[1]{ Phys.\ Lett.\ {\bf #1}}
\newcommand{\PR}[1]{Phys.\ Rev.\ {\bf #1}}
\newcommand{\AN}[1]{Ann. Phys. NY {\bf #1}}
\newcommand{\NP}[1]{ Nucl.\ Phys.\ {\bf #1}}

\begin{document}

\title{Hidden gauge formalism for the radiative decays of
 axial-vector mesons}

\author{
H. Nagahiro$^1$, L.~Roca$^2$, A.~Hosaka$^1$ and E.~Oset$^{1,3}$\\
{\small{\it $^1$Research Center for Nuclear Physics (RCNP),
 Ibaraki, Osaka  567-0047, Japan}}\\
{\small{\it $^2$Departamento de F\'{\i}sica, Universidad de Murcia, E-30071
Murcia,  Spain}}\\
{\small{\it $^3$Departamento de F\'{\i}sica Te\'orica and IFIC,
Centro Mixto Universidad de Valencia-CSIC,}}\\
{\small{\it Institutos de
Investigaci\'on de Paterna, Aptdo. 22085, 46071 Valencia, Spain}}
}

\date{\today}

\maketitle

 \begin{abstract} 
 
 The radiative decay of the axial-vector resonances  into a
pseudoscalar meson and a photon  is studied using the vector meson
Lagrangian obtained from the hidden gauge symmetry (HGS) formalism.
The formalism is well suited to study this problem since it deals
with pseudoscalar and vector mesons in a unified way, respecting
chiral invariance. We show explicitly  the gauge invariance of the
set of diagrams that appear  in the approach and evaluate  the
radiative decay width of the $a_1(1260)$ and $b_1(1235)$  axial
vector meson resonances into $\pi \gamma$. We also include the 
contribution of loops involving anomalous couplings and compare the
results to those obtained previously within another formalism.

\end{abstract}

\section{Introduction}

The radiative decay of mesons has been traditionally advocated as one of the
observables most suited to learn about their nature on which there is a
permanent debate \cite{Kalashnikova:2005zz,Wang:2006mf}.
 Radiative decay of vector mesons 
has been addressed from different points of views 
\cite{Palomar:2003rb,Markushin:2000fa,Oller:2003vf,
Oller:1997yg,Escribano:2008xc}.
The radiative decay of scalar mesons has had a comparatively larger attention.
The radiative decay of the light scalars, $f_0(980)$, $a_0(980)$ has been
studied in 
\cite{Kalashnikova:2005zz,Palomar:2001vg,Black:2002ek,Nagahiro:2008bn,
Escribano:2008xc,Branz:2008ha} 
and the particular case of 
the charmed scalar meson $D_{s0}(2317)$ has been thoroughly studied in 
\cite{Dstar,Lutz:2007sk,Faessler:2007gv}. The axial vector
mesons have also been the subject of study in
\cite{Roca:2006am,gutsche1,gutsche2,Nagahiro:2008zz,Lutz:2008km} from the  perspective
that they are dynamically generated states from the vector-pseudoscalar
interaction \cite{Lutz:2003fm,Roca:2005nm}, or in other words molecular states. 

The idea that the low lying axial vector mesons, like the 
$a_1(1260)$ and $b_1(1235)$, are actually composite particles of a vector and a
pseudoscalar in coupled channels has nontrivial repercussions since one can now
evaluate properties of these resonances as well as determine production cross 
sections and partial decay widths. It has also led to surprising results, since
it was found in \cite{Roca:2005nm} that the formalism produces two $K_1(1270)$
states instead of just one, as commonly assumed, for which strong experimental
support has been found in \cite{Geng:2006yb}
 (see also the PDG \cite{Amsler:2008zz} in this
respect). The evaluation of the radiative decay of the axial vector meson 
resonances into $\gamma$-pseudoscalar meson is a natural test of the theory and
this is the idea behind the work done in 
\cite{Roca:2006am,Nagahiro:2008zz,gutsche1,gutsche2,Lutz:2008km}.
 There are some
differences between these works. In \cite{Roca:2006am,Nagahiro:2008zz} a formalism
involving the vector representation for the vector mesons is employed, 
and approximations used in 
\cite{Lutz:2003fm,Roca:2005nm} are also invoked which render finite the 
results in the calculation of 
the loop functions involved. In \cite{gutsche1,gutsche2} the finiteness
of the results is guaranteed by the use of spatial wave functions for the
molecules. In \cite{Lutz:2008km} a novelty is introduced using 
a tensor representation for the vector mesons, as a consequence of which
the loops involved develop quadratic divergences. These are assumed to be
exactly canceled by some tadpole terms which are not explicitly evaluated.
The tensor formalism for vector mesons was also used in \cite{marco} in the
radiative decay of vector mesons, where the diagrams were found convergent
assuming vector meson dominance, and logarithmically divergent removing this
requirement.

The implementation of a consistent scheme that leads to finite results without
making strong assumptions is most desirable. In that sense, the formalism of
hidden gauge for the vector mesons \cite{Bando:1984ej,Bando:1987br} looks an 
ideal tool, since it deals simultaneously with vector mesons and pseudoscalars,
implements naturally chiral symmetry, leads to the same lowest order chiral
Lagrangian of \cite{Gasser:1984gg} for the pseudoscalar mesons and allows a
 consistent simultaneous treatment
 of vector mesons, pseudoscalars and photon. This latter point is the main issue
in the problem of radiative axial vector meson decays. 

  Another appealing feature of the hidden gauge formalism is that it was proved
in \cite{Ecker:1989yg,Ecker:1988te} that this formalism is equivalent to using
the tensor formalism for the vector mesons, and one can benefit from the
simplicity of the vector formalism, most welcome when dealing with complicated
problems. The hidden gauge formalism also offers the interaction of vector
mesons with pseudoscalars and most importantly, of vector mesons with themselves
for which no Lagrangians are available in the formalism of 
\cite{Ecker:1988te}. 

   Since the axial vector meson resonances are considered here as
composite particles of a vector and a pseudoscalar, the coupling of a
photon is made to the components, and proceeds through loop diagrams
involving the corresponding  vector and pseudoscalar mesons of each
channel.  This is the main framework and provides the largest
contribution. Yet, in some cases where particular large cancellations
appear, it was found in \cite{Nagahiro:2008mn} that contributions of terms
involving anomalous couplings and extra vectors in the loops may be
relevant.  We shall also take this into account.  We shall prove that
the formalism we use, involving  one vector and one pseudoscalar in
the loops, provides finite results for the radiative decay width. The terms
involving the anomalous couplings have logarithmic divergences which
can be cured with a natural cut off or otherwise be related to the
analogous loops appearing in the scattering problem of a vector meson
with a pseudoscalar, leading to similar results in both cases. The
approach presented here leads to a systematic and reliable way to
evaluate radiative decay widths of axial vector mesons we shall  compare
the formalism and the results with those of the former formalism used
in \cite{Roca:2006am,Nagahiro:2008zz}, where couplings of photons to
pseudoscalar and vector mesons are implemented using minimal 
coupling.

\section{The hidden gauge formalism}

The HGS formalism to deal with vector mesons
\cite{Bando:1984ej,Bando:1987br} is a useful and internally
consistent scheme which preserves chiral symmetry. In this
formalism the vector meson fields are gauge bosons of a hidden local
symmetry transforming inhomogeneously. After taking the unitary 
gauge, the vector meson fields transform exactly in the manner as 
in the non linear realization of chiral symmetry \cite{Weinberg:1968de}. 
In Ref.~\cite{Ecker:1989yg} this formalism is
found equivalent to the use of the tensor formalism of 
\cite{Ecker:1988te}, where the vectors transform homogeneously 
under a non-linear realization of chiral symmetry, with the use of
couplings implied in the vector meson dominance formalism (VMD) 
of
\cite{sakurai}. (For a review on the different ways to implement
vector mesons into effective chiral Lagrangians see
Ref.~\cite{Birse:1996hd}).

 Following Ref.~\cite{Ecker:1989yg} 
the Lagrangian involving pseudoscalar mesons, photons and
vector mesons can be written as

\be
{\cal L}={\cal L}^{(2)}+ {\cal L}_{III}
\ee
with
\ba
 {\cal L}^{(2)}=\frac{1}{4}f^2\langle D_\mu U D^\mu U^\dagger+
 \chi U^\dagger+\chi^\dagger U\rangle \\
 {\cal L}_{III}=-\frac{1}{4}\langle V_{\mu\nu}V^{\mu\nu}\rangle
 +\frac{1}{2}M_V^2\langle [V_\mu-\frac{i}{g}\Gamma_\mu]^2\rangle , 
 \label{eq:LIII}
 \ea
where $\langle ...\rangle$ represents a trace over $SU(3)$ matrices. The
covariant derivative is defined by
\be
D_\mu U=\partial_\mu U-ieQA_\mu U+ieUQA_\mu,
\ee
with $Q=diag(2,-1,-1)/3$, $e=-|e|$ the electron charge, and $A_\mu$
the photon field.
The chiral matrix $U$ is given by
\be
U=e^{i\sqrt{2}\phi/f}
\ee
with $f$ the pion decay constant ($f=93\mev$). The $\phi$ and
$V_\mu$ matrices are
the usual $SU(3)$ matrices containing the pseudoscalar mesons and
vector mesons respectively
\be
\phi \equiv 
\left(\begin{array}{ccc}
 \frac{1}{\sqrt{2}} \pi^0 + \frac{1}{\sqrt{6}}\eta_8 
 & \pi^+ & K^+\\
\pi^-&- \frac{1}{\sqrt{2}} \pi^0 + \frac{1}{\sqrt{6}}\eta_8 
& K^0\\
K^-& \bar{K}^0 & -\frac{2}{\sqrt{6}}\eta_8
\end{array}
\right) 
, \,\,
V_\mu \equiv  \left(\begin{array}{ccc} 
\frac{1}{\sqrt{2}} \rho^0 + \frac{1}{\sqrt{2}}\omega 
 & \rho^+ & K^{*+}\\
\rho^-& - \frac{1}{\sqrt{2}} \rho^0 + \frac{1}{\sqrt{2}}\omega 
& K^{*0}\\
K^{*-}& \bar{K}^{*0} & \phi
\end{array}
\right)_{\mu} .
\label{eq:PVmatrices}
\ee
The terms with $\chi$ in ${\cal L}^{(2)}$ provide the mass term for the
pseudoscalars.
For four pseudoscalar meson fields the ${\cal L}^{(2)}$ Lagrangian provides the
well known chiral Lagrangian at lowest order
\be
\widetilde{\cal L}^{(2)}=\frac{1}{12f^2}
\langle [\phi,\partial_\mu \phi]^2 + M\phi^4 \rangle
\label{eq:LPPPP}
\ee
with $M=diag(m_\pi^2,m_\pi^2,2m_K^2-m_\pi^2)$. For the coupling
between two pseudoscalars and one photon the Lagrangian ${\cal L}^{(2)}$
 provides
\be
{\cal L}_{\gamma PP}=-ieA^\mu
\langle Q[\phi,\partial_\mu \phi]\rangle,
\label{eq:LgPP}
\ee
which in this formalism will get canceled with an extra term
coming from ${\cal L}_{III}$, such that ultimately the photon
couples to the pseudoscalars via vector meson exchange, the basic
feature of VMD.

In  ${\cal L}_{III}$, $V_{\mu\nu}$ is defined as 
\be V_{\mu\nu}=\partial_\mu
V_\nu-\partial_\nu V_\mu-ig[V_\mu,V_\nu]
\ee
 and 
\be
\Gamma_\mu=\frac{1}{2}[u^\dagger(\partial_\mu-ieQA_\mu)u
+u(\partial_\mu-ieQA_\mu)u^\dagger]
\ee
with $u^2=U$. The hidden gauge coupling constant $g$ is related to $f$ and the
vector meson mass ($M_V$)
through
\be 
g=\frac{M_V}{2f},
\ee
which is one of the forms of the KSFR relation \cite{ksfr}.
Other properties of $g$ inherent to the VMD
formalism, relating to the tensor formalism of \cite{Ecker:1989yg}
are 
\be \frac{F_V}{M_V}=\frac{1}{\sqrt{2}g}\qquad,\qquad
\frac{G_V}{M_V}=\frac{1}{2\sqrt{2}g}\qquad,\qquad
F_V=\sqrt{2}f\qquad,\qquad G_V=\frac{f}{\sqrt{2}}.
\label{eq:VMDrelations}
\ee

Upon expansion of $[V_\mu-\frac{i}{g}\Gamma_\mu]^2$ up to two
pseudoscalar fields, we find

\ba
[V_\mu-\frac{i}{g}\Gamma_\mu]^2&=&\left(V_\mu-\frac{e}{g}QA_\mu
-\frac{1}{g}\frac{1}{2f^2}\phi e Q A_\mu \phi
+\frac{1}{g}\frac{1}{4f^2}\phi^2e Q A_\mu \right.\nn\\
&+&\left.\frac{1}{g}\frac{1}{4f^2}e Q A_\mu\phi^2
-\frac{i}{g}\frac{1}{4f^2}[\phi,\partial_\mu \phi]\right)^2
\ea
from where we obtain the following interaction Lagrangians
among pseudoscalars ($P$), photons ($\gamma$) and vector mesons
($V$):

\ba
{\cal L}_{V\gamma}&=&-M_V^2\frac{e}{g} A_\mu\langle V^\mu Q\rangle \\
{\cal L}_{V\gamma PP}&=&e \frac{M_V^2}{4g f^2}A_\mu\langle V^\mu 
( Q\phi^2 +\phi^2 Q-2\phi  Q \phi)\rangle \\
{\cal L}_{VPP}&=&-i\frac{M_V^2}{4g f^2}
\langle V^\mu [\phi,\partial_\mu \phi]\rangle 
\label{eq:LVPP}\\
{\cal L}_{\gamma PP}&=&ie A_\mu
\langle Q [\phi,\partial_\mu \phi]\rangle 
\label{eq:LgPP2}\\
\widetilde{\cal L}_{PPPP}&=&-\frac{1}{8f^2}
\langle [\phi,\partial_\mu \phi]^2\rangle . 
\label{eq:LPPPP2}
\ea
The term in Eq.~(\ref{eq:LgPP2}) cancels exactly the term in 
Eq.~(\ref{eq:LgPP}), as mentioned above.
On the other hand, the term of Eq.~(\ref{eq:LPPPP2}) has the same
structure as the derivative term of Eq.~(\ref{eq:LPPPP}) and it is
a most unpleasant term, since added to $\widetilde{\cal L}^{(2)}$
of Eq.~(\ref{eq:LPPPP}) would break the chiral symmetry of the
chiral Lagrangian. However, this term is canceled by the
exchange of vector mesons between the pseudoscalars that result
from the Lagrangian of Eq.~(\ref{eq:LVPP}), ${\cal L}_{VPP}$,
in the limit of $q^2/M_V^2 \to 0$, where $q$ is the momentum carried 
by the exchanged vector meson.
This was already noticed in Ref.~\cite{Weinberg:1968de}.

Furthermore, from the $\langle V_{\mu\nu} V^{\mu\nu}\rangle$ term
of ${\cal L}_{III}$, (see Eq.~(\ref{eq:LIII})), we
obtain the coupling of three vector mesons which is also essential
in the present work
\be
{\cal L}_{VVV}=ig\langle (\partial_\mu V_\nu-\partial_\nu
V_\mu)V^\mu V^\nu\rangle.
\ee

We shall explain the formalism in detail for the $\rho\pi$
component of the $a_1^+$ decay. For this purpose we show in the
Appendix the relevant couplings that allow us to construct the
amplitudes for the radiative decay of the $a_1^+$.

\section{The $VP\to VP$ interaction} 

\begin{figure}[hbt]
\epsfxsize=9cm
\centerline{
\epsfbox{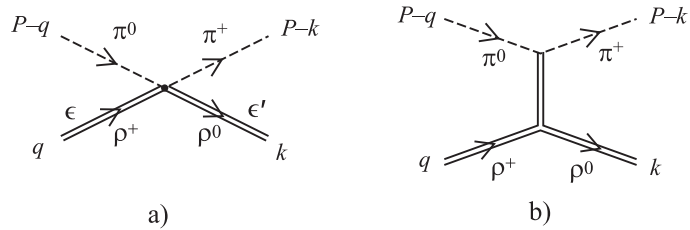}}
\caption{a) Contact interaction from the Lagrangian of Eq.~(\ref{eq:LVPVP}) for
the $\rho^+\pi^0\to\rho^0\pi^+$; 
b) Corresponding diagram provided by the HGS formalism.}
 \label{fig:rhopirhopi}
\end{figure}

In the construction of the interaction kernel for the
vector-pseudoscalar meson interaction, which is used in 
Refs.~\cite{Lutz:2003fm,Roca:2005nm} to generate dynamically 
the axial-vector resonances, the following chiral
Lagrangian is utilized:

\be
{\cal L}=-\frac{1}{4f^2}\langle
[V^\mu,\partial^\nu V_\mu][\phi,\partial_\nu \phi]\rangle,
\label{eq:LVPVP}
\ee
which, for the $\rho^+\pi^0\to\rho^0\pi^+$ gives
\be
{\cal
L}_{\rho^+\pi^0\to\rho^0\pi^+}=\frac{1}{2f^2}(2P-q-k)\cdot(k+q)
\epsilon\cdot\epsilon'
\label{eq:Lrhopirhopi}
\ee
with the assignment of momenta given in Fig.~\ref{fig:rhopirhopi}a)

The HGS formalism leads to the diagram  
of Fig.~\ref{fig:rhopirhopi}b), which can be readily evaluated
 and approximated
using the Feynman rules of the Appendix.

\ba
&&{\cal L}^{HGS}_{\rho^+\pi^0\to\rho^0\pi^+}\simeq
\frac{1}{2f^2}[(k+q)\cdot(2P-k-q)\epsilon\cdot\epsilon' \nn \\
&&-(2k-q)\cdot \epsilon ~(2P-k-q)\cdot\epsilon'
-(2q-k)\cdot \epsilon' ~(2P-k-q)\cdot\epsilon] , 
\label{eq:LHGrhopirhopi}
\ea
where the intermediate $\rho$ propagator $[(q-k)^2-M_V^2]^{-1}$ has
been approximated by $-M_V^{-2}$. As we can see, the first term of
Eq.~(\ref{eq:LHGrhopirhopi}) coincides with the result of the chiral
Lagrangian of Eq.~(\ref{eq:Lrhopirhopi}). The second and third
terms of Eq.~(\ref{eq:LHGrhopirhopi}) are small for small kinetic
energies of the particles since the zeroth component of the
polarization vectors tends to zero as the three-momentum of the vector 
meson goes to
zero to satisfy the Lorenz condition $\epsilon\cdot q=0$. Under
these conditions, the HGS formalism and the chiral Lagrangian of 
Eq.~(\ref{eq:LVPVP}) provide the same vector-pseudoscalar meson
interaction.

The kernel to be used in the Bethe-Salpeter equation is defined as

\be
\widetilde{V}'_{\rho^+\pi^0\to\rho^0\pi^+}\epsilon\cdot\epsilon'
=-{\cal L}_{\rho^+\pi^0\to\rho^0\pi^+}
\ee
which upon projection in isospin $I=1$, for the case of the $a_1^+$
resonance leads to
\be
\widetilde{V}^{(I=1)}_{\rho\pi\to\rho\pi}\epsilon\cdot\epsilon'
=-\widetilde{V}'_{\rho^+\pi^0\to\rho^0\pi^+}\epsilon\cdot\epsilon'
\simeq\frac{1}{2f^2}(k+q)\cdot(2P-k-q)\epsilon\cdot\epsilon'
\label{eq:Vp0}
\ee

\begin{figure}[hbt]
\epsfxsize=12cm
\centerline{
\epsfbox{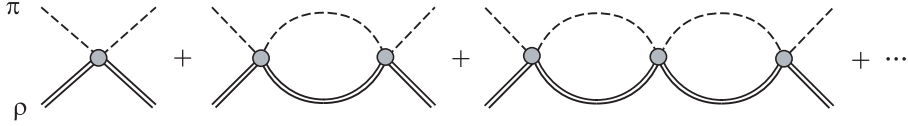}}
\caption{Diagrammatic series of the Bethe-Salpeter equation for the
$\rho\pi$ interaction.}
 \label{fig:BS}
\end{figure}

In Ref.~\cite{Roca:2005nm} the spatial part
$\vec\epsilon\cdot\vec\epsilon\,'$ of the $I=1$ potential in s-wave 
was iterated in the Bethe-Salpeter equation, summing the diagrams of
Fig.~\ref{fig:BS}.
The sum is done in Ref.~\cite{Roca:2005nm} where the following
scattering matrix is obtained:
\be
T\simeq \frac{-\widetilde{V}}{1+\widetilde{V}G}\vec\epsilon\cdot\vec\epsilon\,'
\ee
neglecting terms of $\rho$ momenta
over the mass squared which are very small, where $G$ is the loop
function of a $\rho$ and a $\pi$ conveniently 
regularized \cite{Roca:2005nm} corresponding to 
\be G(P)=\int\frac{i\, d^4
q}{(2\pi)^4}\frac{1}{q^2-m_l^2+i\epsilon}\frac{1}{(P-q)^2-M_l^2+i\epsilon}
\label{eq:gfunction}
\ee

The poles of the $T$-matrix, corresponding to the $a_1^+$ resonance
require $1+\widetilde{V}G=0$ in the second Riemann sheet of the complex energy
plane. In the real axis for the energy of the resonance we will
have
\be
\widetilde{V}G\sim -1\qquad\Longrightarrow\qquad G\sim-\widetilde{V}^{-1}.
\label{eq:VG1}
\ee
This result is approximate because we go from the complex pole
position to the real axis and also because in 
Ref.~\cite{Roca:2005nm} the poles are searched solving the
Bethe-Salpeter equations in coupled channels.
However, since the coupling of $a_1^+$ to $\rho\pi$ is by far the
largest \cite{Roca:2005nm}, the result of
 Eq.~(\ref{eq:VG1}) is a rough approximation which will be used
 later on only for illustrative purposes.
 
\section{Test of gauge invariance}

 Amplitudes which involve a photon must be computed in a way
consistent with the gauge invariance.
As a matter of fact, if all necessary diagrams are properly
taken into account the gauge symmetry is satisfied.
In some cases, however, its proof
is not a trivial matter, especially when higher order loops are
included or approximations are used.
This happens in the present calculation, and therefore, we would
like to discuss it in some detail.

\begin{figure}[hbt]
\epsfxsize=10cm
\centerline{
\epsfbox{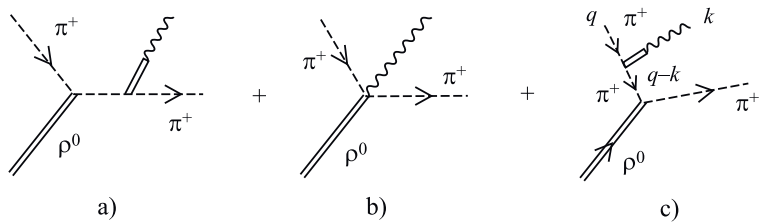}}
\caption{Set of Feynman diagrams which is gauge invariant 
for $\rho^0 \to \pi^+ \pi^- \gamma$.}
 \label{fig:gaugeinv3}
 \end{figure}

\begin{figure}[hbt]
\epsfxsize=10cm
\centerline{
\epsfbox{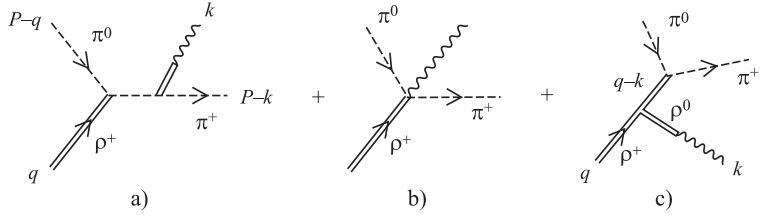}}
\caption{Set of Feynman diagrams which is gauge invariant 
for $\rho^+ \to \pi^+ \pi^0 \gamma$.}
\label{fig:gaugeinv4}
\end{figure}

Let us start with a brief look at a simple
process of physical decay, $\rho \to \pi \pi \gamma$.
This will be used later on to prove
the gauge invariance of the diagrams involved in the axial-vector
meson radiative decay. 
By using the Feynman rules of the Appendix, it is immediate to prove
the gauge invariance of the set of diagrams shown
in Fig.~\ref{fig:gaugeinv4}, upon summing the three diagrams and
substituting $\epsilon^{(\gamma)}\to k$.
This will be used later on to prove
the gauge invariance of the diagrams involved in the axial-vector
meson radiative decay.
Independently, the set of diagrams of  Fig.~\ref{fig:gaugeinv3}
is also gauge invariant.

The test of gauge invariance of the two sets succeeds when all
the external particles are on shell. More concretely, in diagram
c) of Fig.~\ref{fig:gaugeinv4} the intermediate $\rho$ propagator
is
\be
\frac{1}{(q-k)^2-M_V^2}=\frac{1}{q^2-M_V^2-2qk} \to \frac{1}{-2qk}
\label{eq:pipro1}
\ee
In diagram
c) of Fig.~\ref{fig:gaugeinv3} the same occurs with the
intermediate pion propagator. We must keep this in mind since when
the $\rho^+$ in Fig.~\ref{fig:gaugeinv4}c) or the initial
$\pi^+$ in 
Fig.~\ref{fig:gaugeinv3}c) are put inside a loop, as will be the
case in the radiative decay, some extra diagram will be 
demanded to fulfill gauge invariance.
 

\begin{figure}[hbt]
\epsfxsize=12cm
\centerline{
\epsfbox{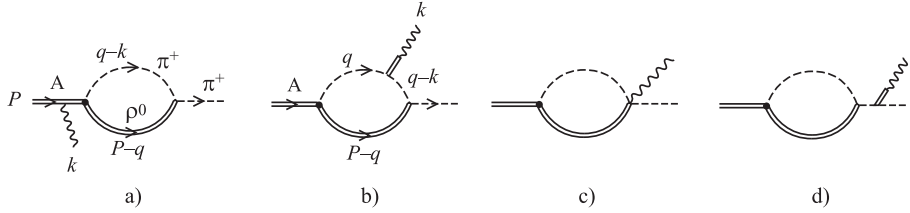}}
\caption{Gauge invariant set of diagrams for the radiative decay
of the axial-vector meson.}
 \label{fig:gaugeinvB1}
\end{figure}

\begin{figure}[hbt]
\epsfxsize=12cm
\centerline{
\epsfbox{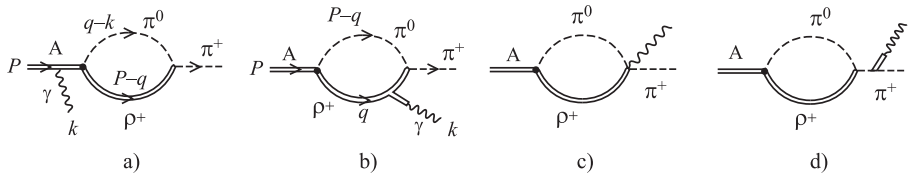}}
\caption{Another set of gauge invariant diagrams for 
the radiative decay
of the axial-vector meson. }
 \label{fig:gaugeinvB2}
\end{figure}

We shall continue considering the $\rho\pi$ channel, the most
important of the $a_1$, for illustrative purposes, although in 
the
final calculations we will consider  the contribution of
all the $VP$ coupled channels.
Following  Ref.~\cite{Roca:2006am} the radiative decay of the
axial-vector mesons is obtained by coupling the photon to its meson
components, which requires the knowledge of the coupling of the
resonance to the different  vector-pseudoscalar components. This
coupling is of the type

\be
V_{a_1^+\rho^+\pi^0}=g_{a_1^+\rho^+\pi^0} \epsilon_A\cdot\epsilon
\ee
 with $\epsilon_A$, $\epsilon$, the polarization vectors of the
 axial and the vector mesons. The couplings $g_i$ are obtained 
in Ref.~\cite{Roca:2005nm} from the residues at the pole positions of the scattering
amplitudes. The set of diagrams needed
for the calculation are given in
Figs.~\ref{fig:gaugeinvB1} and \ref{fig:gaugeinvB2}
 
 
 The connection of this formalism to details of the dynamical
 generation of the axial-vector mesons is discussed in 
 Ref.~\cite{Roca:2006am} and in the analogous case of dynamically
 generated baryons in Ref.~\cite{Doring:2006ub}.
 The gauge invariance of the set of diagrams in 
Figs.~\ref{fig:gaugeinv3} and \ref{fig:gaugeinv4}
would imply the gauge invariance of the set of diagrams b), c), d) 
in Figs.~\ref{fig:gaugeinvB1} and 
 \ref{fig:gaugeinvB2} if the $q$ lines
 were on the mass-shell.
Since this is obviously not the case because they
belong to a loop,
the diagrams a) of Figs.~\ref{fig:gaugeinvB1} and 
\ref{fig:gaugeinvB2} are demanded in order to still fulfill
 gauge invariance \cite{Faessler:2007gv} because they
cancel the effect of the 
 off-shellness of the $q$ line in diagrams 
 \ref{fig:gaugeinvB1}b) and 
 \ref{fig:gaugeinvB2}b). 
 
 Indeed the pion propagator
 with momentum $q-k$ in Fig.~\ref{fig:gaugeinvB1}b is
  \ba
&& \frac{1}{(q-k)^2-m_\pi^2}= \frac{1}{q^2-m_\pi^2-2kq}\nn \\
&&=\frac{1}{-2kq}+\left(\frac{1}{q^2-m_\pi^2-2kq}-
\frac{1}{-2kq}\right) \nn \\
&&=\frac{1}{-2kq}+\frac{1}{2kq}\,\frac{q^2-m_\pi^2}{(q-k)^2-m_\pi^2}.
 \label{eq:pipro2}
 \ea
The first term in  Eq.~(\ref{eq:pipro2}) corresponds to the
propagator of Eq.~(\ref{eq:pipro1}) assuming $q^2=m_\pi^2$ (on
shell pion) and guarantees the cancellation of the last three
diagrams of Fig.~\ref{fig:gaugeinvB1}. The remnant term in 
  Eq.~(\ref{eq:pipro2}) kills the propagator with momentum
  $q$, leaving a loop with just two propagators with momentum $q-k$
  (pion) and $P-q$ (vector), which has then the same topology as
  the diagram of Fig.~\ref{fig:gaugeinvB1}a). It is direct to
  see that with the following coupling of the photon 
  to the axial-vector with
  positive charge in Fig.~\ref{fig:gaugeinvB1}a,
    \be
  -it_{A^+A^+\gamma}=-ie (P+P-k)_\mu \epsilon^{(\gamma)\mu}
  \epsilon_A\cdot\epsilon_A',
  \label{AAgam}
  \ee
  the cancellation of Fig.~\ref{fig:gaugeinvB1}a with the
off-shell part of 
 Fig.~\ref{fig:gaugeinvB1}b, taking the second term of the pion
 propagator of Eq.~(\ref{eq:pipro2}), is exact and one has a gauge
 invariant set of diagrams. A similar reasoning can be made to
show the gauge invariance of the set of Fig.~\ref{fig:gaugeinvB2}.
The Lagrangian of Eq. (\ref{AAgam}) could be obtained via minimum 
coupling neglecting terms of the same order as those neglected 
to convert the $VP \to VP$ interaction of the HGS in the one from the 
effective Lagrangian in section 3. Its use is demanded for 
consistency with the vertex chosen for the $VP \to VP$ interaction.

 \section{The radiative decay of the $a_1^+$ in the hidden gauge
formalism}

Although the diagrams $a, d$, of
Figs.~\ref{fig:gaugeinvB1} and \ref{fig:gaugeinvB2} are needed for
the gauge invariance test, they give null contribution to
the radiative decay amplitude for different reasons,
\begin{itemize}

\item diagram $a$: because of the requirement that
the longitudinal component of the axial-vector propagator does not
develop a pole of the pseudoscalar, which demands that the loop of
Figs.~\ref{fig:gaugeinvB1}a, \ref{fig:gaugeinvB2}a,  vanishes for the
external pion on shell as is the case here
\cite{Kaloshin:1996kz,Roca:2006am}.

\item diagram $d$: because of the Lorenz condition of the axial-vector meson
$P\cdot\epsilon_A=0$ \cite{Faessler:2007gv,Gamermann:2007bm}.

\end{itemize}

Therefore, one can perform the computation of the remaining 
 diagrams $b$ and $c$
explicitly, but it is more rewarding to use a well known procedure
which makes use of the gauge invariance of the set and
automatically accounts for large cancellations which occur between
these diagrams. Following \cite{Roca:2006am} we write for the
amplitude $a_1\to\gamma \pi$

 \be
 T=\epsilon_{A\mu}\epsilon_\nu^{(\gamma)} T^{\mu\nu}
 \label{eq:Tgeneral}
 \ee
 where $T^{\mu\nu}$ can be written, by Lorentz covariance, as
 
 \be
T^{\mu\nu}=a\, g^{\mu\nu} + b\, P^\mu P^\nu + c\, P^\mu k^\nu
 +d\, k^\mu P^\nu + e\, k^\mu k^\nu
 \label{eq:Tmunuterms}
\ee
where the coefficients $a, \cdots, e$ are Lorentz scalar functions of $P$ and $k$.
Note that, due to the Lorenz condition, ${\epsilon_A}_\mu P^\mu=0$,
${\epsilon}_\nu^{(\gamma)} k^\nu=0$, all the terms in Eq.~(\ref{eq:Tgeneral})
vanish except for the $a$ and $d$ terms. On the other hand, gauge
invariance implies that $T^{\mu\nu}k_\nu=0$, from where one gets
\be
a=-d\,P\cdot k.
\label{eq:ad}
\ee

This is obviously valid in any reference frame, however, in the
axial-vector meson rest frame and taking the Coulomb gauge for the
photon, only the $a$ term survives in  Eq.~(\ref{eq:Tgeneral})
since  $\vec P=0$ and $\epsilon^0=0$. This means that, in the end,
we will only need the $a$ coefficient for the evaluation of the
process. However, the $a$ coefficient can be evaluated from the $d$
term thanks to Eq.~(\ref{eq:ad}). The advantage of doing this is
that there are few mechanisms contributing to the $d$ term and by
dimensional reasons the number of powers of the loop momentum in
the numerator will be reduced, as will be clearly manifest from the
discussion below.

In the present case it is easy to see that the diagrams $c$ of 
Figs.~\ref{fig:gaugeinvB1} and \ref{fig:gaugeinvB2} do not
contribute to the $d$ coefficient and hence one only has to
evaluate the diagrams $b$ of Figs.~\ref{fig:gaugeinvB1} and
\ref{fig:gaugeinvB2}.
The details on how to evaluate the $d$ coefficient using the
Feynman parametrization of the amplitudes are given in 
Ref.~\cite{Roca:2006am}.
The only difference from the previous case 
is in the diagram Fig.~\ref{fig:gaugeinvB2} b) where the effective
$\gamma VV$ vertex mediated by the vector meson propagator 
through the vector meson dominance has extra terms (see section 7 for 
more details).  
It turns out that the presence of the extra terms which come
from the self-interaction vertex of a non-abelian gauge theory
plays a crucial role to make the $d$-coefficient extracted from 
Fig.~\ref{fig:gaugeinvB2} b) finite.  

The total amplitude for the radiative decay is obtained as 
a sum over the diagrams of Figs.~\ref{fig:gaugeinvB1} and \ref{fig:gaugeinvB2}, 
\be
T=T^{Fig.\ref{fig:gaugeinvB1}} +T^{Fig.\ref{fig:gaugeinvB2}}\, .
\ee
 For illustrative purposes we will consider in the present paper the
same decays as in Ref.~\cite{Roca:2006am}, which are
$a_1^+\to\pi^+\gamma$ and $b_1^+\to\pi^+\gamma$.
For these decays the $K^* K$ channels are also needed. 
The general  expression for the amplitude for the kind of
mechanisms shown in Figs.~\ref{fig:gaugeinvB1}
and \ref{fig:gaugeinvB2} are
\begin{eqnarray}
T^{Fig.\ref{fig:gaugeinvB1}} &=& g'_{AVP} Q c_{VPP} 
\frac{M_V G_V}{\sqrt{2}f^2} P\cdot k\,
\epsilon_A \cdot \epsilon^{(\gamma)} \nonumber\\
&\times&\int_0^1 dx \int_0^x dy
\frac{1}{8\pi^2}
\frac{1}{s+i\varepsilon}
\left((1-x)(2-y)-y(1-x)\frac{m_{\pi^+}^2-m_P^2}{M_V^2}
\right) , 
\label{eq:typeb}
\end{eqnarray}
 where 
 \be
 s=(1-x)(xM_{A}^2-M_V^2-2 y P\cdot k)-xm_P^2,
 \ee
with $m_P$ and $M_V$ the masses of the pseudoscalar and vector mesons in
the loop, $m_{\pi^+}$ the mass of the final state pion and $M_A$ the mass 
of the axial vector meson,
\ba
T^{Fig.\ref{fig:gaugeinvB2}} &=& -g'_{AVP} Q c_{VPP} 
\frac{M_VG_V}{\sqrt{2}f^2} P\cdot k\,
\epsilon_A\cdot\epsilon^{(\gamma)} \nonumber\\
&\times&
\int_0^1 dx \int_0^x dy
\frac{1}{16\pi^2}\frac{1}{s'+i\varepsilon}
\left(5x-2y+xy-y(1-x)\frac{m_{\pi^+}^2-m_P^2}{M_V^2}\right) , 
\label{eq:typec}
 \ea
 where 
 \be
 s'=(1-x)(xM_{A}^2-m_P^2-2 y P\cdot k)-xM_V^2.
 \ee 
In Eqs.~(\ref{eq:typeb}) and (\ref{eq:typec}),
  $g'_{AVP}$ are
the $AVP$ coupling constants in the charge base. These
 coefficients  are related to the
 $g_{AVP}$
 in isospin base, obtained in
Ref.~\cite{Roca:2005nm},  through the transformation
\begin{equation}
g'_{AVP} = {\cal C} \times g_{AVP},
\label{eq:gAVP}
\end{equation}
where ${\cal C}$ are coefficients which depend on the different 
$AVP$ channels. We use the values of $g_{AVP}$ obtained in
Refs.~\cite{Roca:2005nm,Roca:2006am} by evaluating the residua at
the pole position of the different $VP\to VP$ scattering
amplitudes.
In the previous equations, ${\cal C}$,
 $Q$ and $c_{VPP}$ are coefficients\footnote{
 Note the different sign in the definition of $e$ with respect to
 Refs.~\cite{Roca:2006am,Nagahiro:2008zz}, since $e$ is taken negative in
the hidden gauge formalism.}
given in tables~\ref{tab:a1coeff} and \ref{tab:b1coeff}.

\begin{table}[h]
\begin{center}
\begin{tabular}{rc|r|r|r} 
\multicolumn{2}{r|}{$a_1^+(1260)\rightarrow\pi^+\gamma$} & ${\cal C}$ & $Q$
 & $c_{VPP}$ \\ \hline\hline 
 type Fig.\ref{fig:gaugeinvB1} & $\overline{K^*}^0K^+$ &
 $1/\sqrt{2}$ & $-e$ & $1$ \\
 & $\rho^0\pi^+$ & $1/\sqrt{2}$ & $-e$ & $-\sqrt{2}$ \\\hline
 type Fig.\ref{fig:gaugeinvB2} & ${K^*}^+\overline{K^0}$ & 
 $-1/\sqrt{2}$ & $-e$ & $-1$ \\
 & $\rho^+\pi^0$ & $-1/\sqrt{2}$ & $-e$ & $\sqrt{2}$ \\
\end{tabular}
\end{center}
\caption{Coefficients for $a_1^+(1260)\rightarrow\pi^+\gamma$ decay.}
\label{tab:a1coeff}
\end{table}

\begin{table}[h]
\begin{center}
\begin{tabular}{rc|r|r|r} 
\multicolumn{2}{r|}{$b_1^+(1235)\rightarrow\pi^+\gamma$} & ${\cal C}$ & $Q$
 & $c_{VPP}$ \\ \hline\hline 
 type Fig.\ref{fig:gaugeinvB1} & $\overline{K^*}^0K^+$ &
 $1/\sqrt{2}$ & $-e$ & $1$  \\\hline
 type Fig.\ref{fig:gaugeinvB2} & ${K^*}^+\overline{K^0}$ & 
 $1/\sqrt{2}$ & $-e$ & $-1$  \\
\end{tabular}
\end{center}
\caption{Coefficients for $b_1^+(1235)\rightarrow\pi^+\gamma$ decay.}
\label{tab:b1coeff}
\end{table}

 \section{Comparison of the results with the tree level}
 \label{sec:comptree}
 
One is now asked to address the contribution of the tree level
diagram of Fig.~\ref{fig:tree}.
Using the Feynman rules of the Appendix, one finds
\be
 T^{tree}=g_{a_1\rho^0\pi^+}\frac{e}{\sqrt{2}g}
 \epsilon_A\cdot\epsilon^{(\gamma)}\qquad;\qquad 
 g_{a_1\rho^0\pi^+}=- g_{a_1\rho^+\pi^0}
 \label{eq:tree}
\ee
However, this term should not be added in the hidden gauge
formalism since it would lead to doublecounting. Indeed, we are
going to prove that this term is identical to the diagram of 
Fig.~\ref{fig:gaugeinvB2}b and thus it has already been counted. This
observation was rightly stated in Ref.~\cite{Lutz:2008km}
and it holds for the HGS formalism. It cannot be applied to the
model used in Ref.~\cite{Roca:2006am} where the diagram of 
Fig.~\ref{fig:gaugeinvB2}b is not introduced and instead a direct
coupling of the photon to the vector meson arising from minimal
coupling in the Proca equation was used. We shall come back to this
point later on.

\begin{figure}[hbt]
\epsfxsize=.25\textwidth
\centerline{
\epsfbox{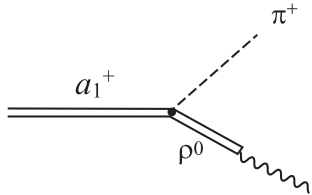}}
\caption{Tree level diagram for $a_1^+\to\gamma\pi^+$ decay.}
 \label{fig:tree}
\end{figure}

\begin{figure}[hbt]
\epsfxsize=.8\textwidth
\centerline{
\epsfbox{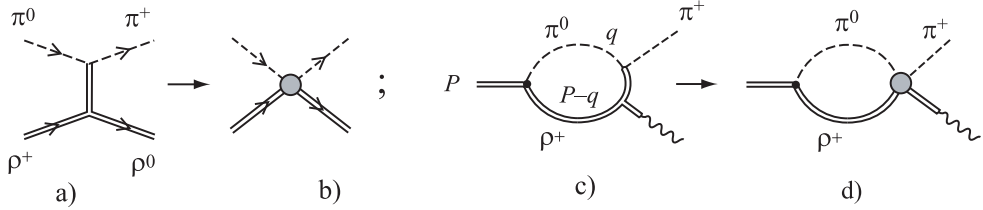}}
\caption{Vector meson exchange in the limit $m_V \to$ large. }
 \label{fig:Vp}
\end{figure}

Let us reinterpret the diagram of Fig.~\ref{fig:gaugeinvB2}b
in terms of the tree level diagram of Fig.~\ref{fig:tree}. 
The essential argument comes 
by comparing the diagram of Fig.~\ref{fig:gaugeinvB2}b with 
the diagrams for the 
dynamical generation of the $a_1$ as shown in Fig.~\ref{fig:BS}.  
The relation to the former can be achieved by taking 
the limit $m_V \to$ large for the vector meson which emits the pion of the
final state, as shown in Fig.~\ref{fig:Vp}.   
Since the heavy vector meson exchange has been used for the construction 
of $a_1$, the diagram of Fig.~\ref{fig:Vp}c, omitting the photon, is
equivalent to the sum over  
diagrams of Fig.~\ref{fig:BS} excluding the first tree diagram.  
Near the resonance pole of $a_1$, however, the first tree diagram 
can be neglected and hence the diagram of Fig.~\ref{fig:Vp}c, omitting
the photon, becomes 
equivalent to Fig.~\ref{fig:BS}.  
This enables one to reinterpret the diagram of 
Fig.~\ref{fig:gaugeinvB2}b (or~\ref{fig:Vp}d) as equivalent to the 
tree diagram of Fig.~\ref{fig:tree}.  

Let us see how this occurs analytically.
For this purpose we factorize the vertex of $VP \to VP$ 
in terms of the potential $\widetilde{V}'$ of Eq.~(\ref{eq:Vp0}).
Then the 
diagram \ref{fig:gaugeinvB2}b now acquires the topology of 
Fig.~\ref{fig:Vp}d and can be computed as
\ba
-iT^{Fig.\ref{fig:Vp}d}
&=&
-ig_{a_1\rho^+\pi^0}\int 
\frac{d^4q}{(2\pi)^4}
\,\epsilon_A\cdot\epsilon
\,\frac{i}{q^2-m_\pi^2}\,\frac{i}{(P-q)^2-M_V^2}
\nonumber \\
&\times&
(-i)\widetilde{V}'_{\rho^+\pi^0\to\rho^0\pi^+}\, 
\epsilon\cdot\epsilon^{(\gamma)}
\frac{i}{M_V^2}(-i)\frac{1}{\sqrt{2}}M_V^2\frac{e}{g} . 
\ea
Summing over the $\epsilon$ polarization, neglecting the 
$q^\mu q^\nu/M_V^2$ terms of the 
$\rho$ propagator and considering Eq. (\ref{eq:Vp0}) and the second of the eqs.
(\ref{eq:tree}), we have
 \ba
T^{Fig.\ref{fig:Vp}d}
&\simeq& 
-\,g_{a_1\rho^+\pi^0}
\frac{e}{\sqrt{2}\,g}
\widetilde{V}'_{\rho^+\pi^0\to\rho^0\pi^+}G
\,\epsilon_A\cdot\epsilon^{(\gamma)}
\nonumber \\
&=&
-\,g_{a_1\rho^0\pi^+}
 \frac{e}{\sqrt{2}\,g}\widetilde{V}^{(I=1)}_{\rho\pi\to\rho\pi}G
 \,\epsilon_A\cdot\epsilon^{(\gamma)}, 
 \label{eq:liketree}
\ea
where $G$ is given by Eq.~(\ref{eq:gfunction}) and we regularize it as done in
Ref.~\cite{Roca:2005nm}. 
Eq. (\ref{eq:liketree}) coincides with 
Eq.~(\ref{eq:tree}) if $\widetilde{V}^{(I=1)}_{\rho\pi\to\rho\pi}G\simeq -1$,
which is the condition at 
the pole position of the axial vector meson, as discussed in section
3.  
However, note that
$\widetilde{V}^{(I=1)}_{\rho\pi\to\rho\pi}G\simeq -1$ is only true at the pole
position and assuming one channel.
Note also that
in reality the
$a_1$ resonance is very wide and hence it is far from the real
axis, and also the effect of the other channels is not negligible. 
Furthermore, in the loop of 
Fig.~\ref{fig:Vp}d, the factorization of  $\widetilde{V}^{(I=1)}_{\rho\pi\to\rho\pi}$
does not hold exactly (although quite accurately).
Nevertheless, an actual exercise
tells us that these two terms are of the same order of magnitude.  
 
The former discussion offers the possibility to evaluate the set of
diagrams of Fig.~\ref{fig:gaugeinvB2} in a different way. We state
that diagram  \ref{fig:gaugeinvB2}b is the tree level of 
Fig.~\ref{fig:tree} and then we must add to it the diagram of 
Fig.~\ref{fig:gaugeinvB2}c. This diagram is also 
evaluated with the same approximations done above and we obtain
\ba
 T^{Fig.\ref{fig:gaugeinvB2}c}=g_{a_1\rho^0\pi^+}
 M_V^2\frac{e}{2\sqrt{2}\,gf^2}
G \,\epsilon_A\cdot\epsilon^{(\gamma)} .
\ea
The ratio of $T^{Fig. \ref {fig:gaugeinvB2}c}$ to the tree level is
estimated as
\ba
\frac{T^{Fig.\ref{fig:gaugeinvB2}c}}{T^{tree}}=G\frac{M_V^2}{2f^2}
\simeq -\frac{M_V^2}{2f^2\widetilde{V}^{(I=1)}_{\rho\pi\to\rho\pi}}
\simeq -\frac{M_\rho^2}{2M_{a_1}^2-2m_\pi^2-2M_\rho^2}\simeq-0.3.
\ea
This means that the whole set of diagrams of 
Fig.~\ref{fig:gaugeinvB2} can be approximated in terms of the
tree level by
\be
T^{Fig.\ref{fig:gaugeinvB2}}\simeq 0.7 T^{tree}.
\label{eq:ratio6tree}
\ee 
 

\section{Comparison with a previous model}
 
\begin{figure}[hbt]
\epsfxsize=11cm
\centerline{
\epsfbox{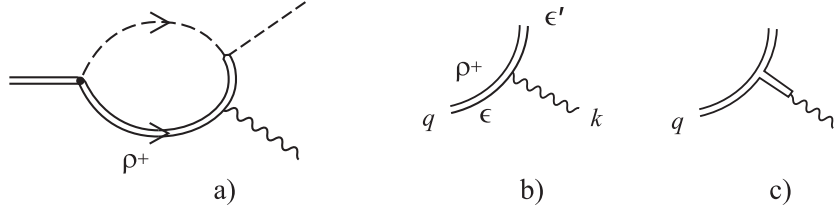}}
\caption{a) and b): diagram and vertex used in the work 
of Ref.~\cite{Roca:2006am}; c): vertex used in the present work.}
 \label{fig:diagold}
\end{figure}

In this section we compare the results obtained in the hidden
gauge formalism with those of the model of 
Ref.~\cite{Roca:2006am}. The model of Ref.~\cite{Roca:2006am}
obtains the couplings via the Proca equation through minimal
coupling. 
The diagrams obtained from the Proca equation corresponding to 
Fig.~\ref{fig:gaugeinvB1} are identical
to those obtained in the previous sections.
 Those of Fig.~\ref{fig:gaugeinvB2} were the
same except diagram \ref{fig:gaugeinvB2}b. This diagram was absent
in the approach of Ref.~\cite{Roca:2006am}. Instead one had the
diagram of Fig.~\ref{fig:diagold}a
 since minimal coupling on the Proca equation leads to a direct
 photon coupling to the vector meson, Fig.~\ref{fig:diagold}b.
  The diagram of 
 Fig.~\ref{fig:diagold}b gives rise in this case to the contribution
\ba
 T^{Fig.\ref{fig:diagold}b}=e[2\epsilon_\alpha^{(\gamma)}q^\alpha
 \epsilon'_\beta\epsilon^\beta
 -\epsilon_\alpha^{(\gamma)}\epsilon^\alpha 
 \epsilon'_\beta q^\beta
 -\epsilon_\alpha^{(\gamma)}\epsilon'^\alpha 
 \epsilon_\beta (q-k)^\beta] , 
 \ea
 while
\ba
 T^{Fig.\ref{fig:diagold}c}=e[
 (2q-k)_\alpha\epsilon^{(\gamma)^\alpha}
 \epsilon_\beta\epsilon'^\beta
 -(q+k)^\beta\epsilon'_\beta\epsilon_\alpha^{(\gamma)}\epsilon^\alpha 
 +\epsilon_\alpha^{(\gamma)}\epsilon'^\alpha 
 \epsilon_\beta (2k-q)^\beta].
 \ea
Note that these two operators are different. As a consequence of
this, the diagram of Fig.~\ref{fig:diagold}a cannot be identified
with the tree level diagram.
 
It is worth looking at the difference between these two operators
\ba
 T^{Fig.\ref{fig:diagold}c}- T^{Fig.\ref{fig:diagold}b}=
 e[k\cdot\epsilon \,\epsilon^{(\gamma)}\cdot\epsilon'
 -k\cdot\epsilon' \,\epsilon^{(\gamma)}\cdot\epsilon].
\ea 
This difference is gauge invariant, which proves that if
in the HGS approach the set of diagrams taken is gauge invariant, so
was the set of diagrams taken in Ref.~\cite{Roca:2006am}.
This was already stated there.
Since, as we have shown, the diagram of 
Fig.~\ref{fig:diagold}a is not equivalent to the tree level,
unlike the diagram of Fig.~\ref{fig:gaugeinvB2}b, the question of
the tree level diagram is reopen.  
Hence, in Ref.~\cite{Roca:2006am} the tree contribution 
was considered, invoking the approach of 
Refs.~\cite{Ecker:1988te,Roca:2003uk} where it appears with
strength similar to the one found here.

To continue with the comparison, let us quote further that 
the sum over the set
of diagrams of Figs.~\ref{fig:gaugeinvB2}a, 
\ref{fig:gaugeinvB2}c, \ref{fig:gaugeinvB2}d
and  \ref{fig:diagold}a 
was found to provide a very small contribution, negligible in practice 
\cite{Roca:2006am,Nagahiro:2008zz}.
As a consequence, for this set of diagrams plus the tree level one
obtains essentially the tree level contribution, while the set of
diagrams of Fig.~\ref{fig:gaugeinvB2} in the HGS formalism has been
found to lead to a contribution about 0.7 times the tree level. The
two formalisms hence lead to a difference of 30\% of the tree level.  
Since the decay width is proportional to the square amplitude, 
its actual value in the HGS formalism is expected to be about 
a half ($0.7^2$) of the decay width from the tree amplitude.  
This rough estimation is indeed consistent with the present 
result as shown in tables \ref{tab:width_a1} and \ref{tab:width_b1}.  

%

\section{New anomalous mechanisms}
\begin{figure}[hbt]
\epsfxsize=5cm
\centerline{
\epsfbox{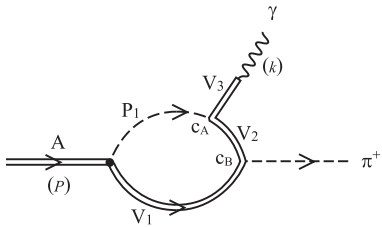}}
\caption{Feynman diagram containing the anomalous vertices.}
 \label{fig:ano}
\end{figure}

Besides those mechanisms considered so far, other diagrams could
provide a relevant contribution to the radiative decay, like those
shown in Fig.~\ref{fig:ano}. 
The main peculiarity of the diagram of Fig.~\ref{fig:ano} is that it contains
two anomalous $VVP$ vertices, which in principle one could expect  
to be small due to the higher order nature of the anomalous term 
in the chiral expansion.  
The $VVP$ interaction is anomalous\cite{wesszumino}
and accounts for a process that
does not conserve intrinsic 
parity\footnote{The 
intrinsic parity of a particle is defined as
follows: it is $+1$ if the particle transforms as a true tensor
of that rank, and $-1$ if it transforms as a pseudotensor, 
{\it e.g.} $\pi$, $\gamma$, $\rho$ and $a_1$ have intrinsic
parity $-1$, $+1$, $+1$ and $-1$ respectively.},
and can be obtained from the gauged
Wess-Zumino term (see {\it e.g.} 
Refs.~\cite{Meissner:1987ge,Pallante:1992qe}).
The expectation of small amplitudes
was the reason why this double-anomalous
mechanisms were not considered in
Refs.~\cite{Roca:2006am,Nagahiro:2008zz}. 
However, later works 
on the radiative decays of scalar mesons~\cite{Nagahiro:2008mn,Nagahiro:2008bn} 
showed the
importance of these mechanisms in radiative decay processes, which 
was also suggested in Ref.~\cite{Lutz:2007sk}.  
The importance of the anomalous process is also shown 
in another context of the kaon photoproduction~\cite{Ozaki:2007ka}.  
In all these examples, as the relevant energy becomes larger the 
role of the anomalous contribution becomes more relevant.  
Therefore we are going to evaluate its contribution in the present
work.


The $VVP$ Lagrangian is
\cite{Bramon:1992kr,Oset:2002sh,Pallante:1992qe}:
\be
{\cal L}_{VVP}=\frac{G'}{\sqrt{2}}\epsilon^{\mu\nu\alpha\beta}
\langle\partial_\mu V_\mu\partial_\alpha V_\beta P\rangle,
\label{eq:LVVP}
\ee
where $G'=3g'^2/(4\pi^2f)$ with $g'=-G_V M_\rho /(\sqrt{2}f^2)$.
Since in the loops of Fig.~\ref{fig:ano} 
we have two vertices of the type $VVP$,
the amplitude is proportional to $G'^2$ or $g'^4$. Hence, the
contributions to the decay width of the loops of Fig.~\ref{fig:ano}
 go like $g'^8$.
Thus the decay width is very sensible to the exact value of the
$VVP$ couplings. 
In order to fine tune the numerical value of the $VVP$
coupling we proceed in a similar way as in
Ref.~\cite{Nagahiro:2008mn}: we normalize the $G'$ coupling
multiplying it by a factor $N_i$ such
that the $V\to P\gamma$ decay widths agree with the experimental
results (see Ref.~\cite{Nagahiro:2008mn} for details and exact 
definition of $N_i$).

The amplitude of the diagram of Fig.~\ref{fig:ano} is logarithmically 
divergent. Following the procedure of  \cite{Nagahiro:2008mn} one can
isolate a divergent  part having a loop structure with a pair of the same two meson
propagators as appearing in the scattering problem (in the present case, a
pseudoscalar and a vector).  
This term is naturally associated with the
$VP$ loop function, $G_{VP}$, of Eq.~(26). 
In  Ref.~\cite{Nagahiro:2008mn}, it was
also shown that 
the logarithmically divergent term can be regularized 
with a cut off of natural size ($\sim 1$ GeV), having lead to very
similar results. 
Thus we obtain:
\ba
T^{\rm anom}&=&
g'_{AP_1V_1} c_A c_B\lambda_{V_3}
|e|\frac{N_AN_B G'^2 F_V}{2 M_{V_3}} 
\epsilon_A\cdot\epsilon^{(\gamma)} P\cdot k\nonumber\\
&\times&\left[\int_0^1 dx \int_0^x dy \frac{1}{16\pi^2}
\frac{1}{s+i\varepsilon}\left((P^2/2-k\cdot P)(1-x)^2
-\frac{m_{V_2}^2}{2}\right) 
+\frac{1}{2}G_{VP}(P^0,m_{V_1},m_{P_1})\right] , 
\nonumber\\
 \label{eq:dbconv}
\ea
where $s=(P^2x-2P\cdot k y) (1-x)-m_{V_1}^2
+(m_{V_1}^2-m_{P_1}^2)x-(m_{V_2}^2-m_{P_1}^2) y$
and $\lambda_{V}$ is $1$, $1/3$, $-\sqrt{2}/3$ for $V=\rho$,
 $\omega$,
$\phi$ respectively.
The coefficietns $c_A$ and $c_B$ are coming
from the $VV'P$ vertex defined as $ c_i VV'P$ after taking the trace in 
Eq.~(\ref{eq:LVVP}) and given in tables  
\ref{tab:coef_anom_a1} and \ref{tab:coef_anom_b1}.

\begin{table}[htbp]
\begin{center}
\begin{tabular}{c|r|r|r} 
\multicolumn{4}{l}{$a_1^+(1260) \rightarrow \pi^+\gamma$} \\\hline\hline
$P_1V_1V_2 V_3$ & ${\cal C}$ & $c_A$ & $c_B$ \\\hline
$\pi^0\rho^+\omega\rho^0$ & $-1/\sqrt{2}$ & $\sqrt{2}$ & $\sqrt{2}$ \\
$K^+ \bar{K}^{*0} K^{*+}\rho^0$ & $1/\sqrt{2}$ & $1/\sqrt{2}$ & 1 \\
$\bar{K}^0K^{*+}\bar{K}^{*0}\rho^0$ & $-1/\sqrt{2}$ & $-1/\sqrt{2}$ & 1 \\
\end{tabular}
\end{center}
\caption{Coefficients for the anomalous term of the
 $a_1^+(1260)\rightarrow\pi^+\gamma$ decay.}
\label{tab:coef_anom_a1}
\end{table}
\begin{table}[htbp]
\begin{center}
\begin{tabular}{c|r|r|r} 
\multicolumn{4}{l}{$b_1^+(1235) \rightarrow \pi^+\gamma$} \\\hline\hline
$P_1V_1V_2 V_3$ & ${\cal C}$ & $c_A$ & $c_B$ \\\hline
$\pi^+\omega \rho^+\omega$ & $-1$ & $\sqrt{2}$ & $\sqrt{2}$ \\
$\eta\rho^+\omega\omega$ & $-1$ & $2/\sqrt{3}$ & $\sqrt{2}$ \\
$K^+ \bar{K}^{*0} K^{*+}\omega$ & $1/\sqrt{2}$ & $1/\sqrt{2}$ & 1 \\
$K^+ \bar{K}^{*0} K^{*+}\phi$ & $1/\sqrt{2}$ & 1 & 1 \\
$\bar{K}^0K^{*+}\bar{K}^{*0}\omega$ & $1/\sqrt{2}$ & $1/\sqrt{2}$ & 1 \\
$\bar{K}^0K^{*+}\bar{K}^{*0}\phi$ & $1/\sqrt{2}$ & 1 & 1 \\
\end{tabular}
\end{center}
\caption{Coefficients for the anomalous term of the
 $b_1^+(1235)\rightarrow\pi^+\gamma$ decay.}
\label{tab:coef_anom_b1}
\end{table}


\section{Numerical results}

 With the amplitudes obtained above, the decay width
for the axial-vector mesons into one pseudoscalar
 meson and one photon is given by
\begin{equation}
\Gamma(M_A) = \frac{|\vec{k}|}{12\pi M_A^2} |T|^2,
\end{equation}
where $M_A$ stands for the mass of the decaying axial-vector meson
and $T$ is the sum of the amplitudes from the loop
mechanisms removing the $\epsilon_A\cdot\epsilon$ factor.
The former expression is valid for narrow axial-vector
resonances.
In order to take into account the finite width of the axial-vector
meson we fold the previous expression
with the mass distribution:
\begin{equation}
\Gamma_{A\rightarrow P\gamma} = -\frac{1}{\pi}
\int_{(M_A-2\Gamma_A)^2}^{(M_A+2\Gamma_A)^2}
ds_A\,
Im
\left\{\frac{1}{s_A-M_A^2+iM_A\Gamma_A}\right\}
\Gamma(\sqrt{s_A})
\Theta(\sqrt{s_A}-\sqrt{s_A^{th}}),
\end{equation}
where $\Theta$ is the step function, $\Gamma_A$ is the total
axial-vector meson width and $s_A^{th}$ is the threshold for the
dominant $A$ decay channels.

Similarly, since the $\rho$ and $K^*$ mesons have relatively large
widths, we have also taken into account the mass distribution of
these states in the loop functions of 
Figs.~\ref{fig:gaugeinvB1} and \ref{fig:gaugeinvB2}. 
This is done by folding 
$T^{Fig.\ref{fig:gaugeinvB1}}$,
$T^{Fig.\ref{fig:gaugeinvB2}}$, with the spectral
function of the $\rho$ and $K^*$:

\begin{equation}
T^{Fig.\ref{fig:gaugeinvB1},Fig.\ref{fig:gaugeinvB2}}\to
T^{Fig.\ref{fig:gaugeinvB1},Fig.\ref{fig:gaugeinvB2}}= -\frac{1}{\pi}
\int_{(M_V-2\Gamma_V)^2}^{(M_V+2\Gamma_V)^2}
ds_V\,
Im
\left\{\frac{1}{s_V-M_V^2+iM_V\Gamma_V}\right\}
T^{Fig.\ref{fig:gaugeinvB1},Fig.\ref{fig:gaugeinvB2}}(\sqrt{s_V}).
\end{equation}
\noindent
The corrections from this source are small, they change the
radiative decay widths at the level of $2$\% or below.

%
In tables \ref{tab:width_a1} and \ref{tab:width_b1}
we can see various
contributions of different kinds of loops to the radiative
decays.
The theoretical
errors have been obtained by doing a Monte-Carlo sampling of
the parameters of the model within their uncertainties, as
explained in Refs.~\cite{Roca:2006am,Nagahiro:2008zz}.
 Due to the strong role played by the interferences between
different mechanisms, as will be explained below, and the
approximations involved in the relations between couplings, like
those in Eq.~(\ref{eq:VMDrelations}), we have multiplied the
errors by two to account safely for the theoretical uncertainties.  
 We also show in the tables the result for the tree level
with the model of Refs.~\cite{Roca:2006am,Nagahiro:2008zz} which must
not be consider if using the HGS formalism as explained in
section~\ref{sec:comptree}. 
In the last row the experimental values provided by the 
PDG \cite{Amsler:2008zz} are given, however these numbers refer to
one single old experiment for each decay. 
From the table, the theoretical value of the total decay width 
for $a_1$ seems underestimated unlike the previous 
results~\cite{Roca:2006am}, while for $b_1$ the agreement 
is good.

If we look at more details of the theoretical values, 
the amplitudes of Figs.~\ref{fig:gaugeinvB1} and 
\ref{fig:gaugeinvB2} are destructively added for the $a_1$ decay,
 and the sum of them is smaller than each contribution.  
 For the $b_1$ case this interference is constructive.
In the present calculation, we have a new contribution 
from the anomalous term which is relatively large 
as compared to the normal contributions of 
Figs.~\ref{fig:gaugeinvB1} and \ref{fig:gaugeinvB2}
for the case of the $a_1$.  
It is also interesting to observe that the anomalous contribution 
is dominated by the $\rho\pi$ loop.  
It is therefore important to consider the anomalous 
terms if they exist.  
For the total theoretical amplitude for $a_1$ decay, 
the anomalous term has an opposite phase
to the sum of contributions from Figs.~\ref{fig:gaugeinvB1} and
\ref{fig:gaugeinvB2}, and 
so the net amplitude and the resulting decay width 
become small, 133 keV as compared with 
640 $\pm$ 246 keV of the experimental value.  
For the case of $b_1$ the normal contributions 
already agree well as compared with experimental data, 
while the anomalous contribution is very small.
Therefore, the total decay width agrees well 
with experimental data.


From the results in the tables it can be seen that 
the loops from 
Fig.~\ref{fig:gaugeinvB2} contribute much more than in the
formalism of Refs.~\cite{Roca:2006am} 
where it was shown to be very small.
This should not be surprising since the loop of
Fig.~\ref{fig:gaugeinvB2}b is very different from the one of
Fig.~\ref{fig:diagold}a, evaluated in Ref.~\cite{Roca:2006am}.
The actual value from the loop of Fig.~\ref{fig:gaugeinvB2} is
consistent with the rough estimation of  
Eq.~(\ref{eq:ratio6tree}); 
for the case of $a_1$, they contribute 373 keV
which is consistent with $0.7^2 \times 647$ keV.   
For the case of the $b_1$ the argument holds only qualitatively;
57 keV of the contribution of Fig.~\ref{fig:gaugeinvB2} which 
is compared with the tree contribution of 67 keV.  
\begin{table}[htbp]
\begin{center}
\begin{tabular}{crr} 
\multicolumn{3}{l}{$a_1^+(1260) \rightarrow \pi^+ \gamma$} \\\hline\hline
tree level with model of Refs.~\cite{Roca:2006am,Nagahiro:2008zz} & $\phi$ & - \\
 & $\omega$ & - \\
 & $\rho$ & 647 \\\cline{2-3}
 & total & 647 \\\hline\hline
loops type Fig.5 & $K^*K$ & 14 \\
 & $\rho\pi$ & 119 \\\cline{2-3}
 & total & 171 \\\hline
loops type Fig.6 & $K^*K$ & 30 \\
 & $\rho\pi$ & 213 \\\cline{2-3}
 & total & 373 \\\hline
\multicolumn{2}{c}{total (Fig.5+Fig.6)} & 103 \\\hline
loops anomalous & $\rho\pi$ & 163 \\
 & $\bar{K}^{*0} K^+$ & 1.4 \\
 & ${K^*}^+ \bar{K}^0$ & 1.4 \\\cline{2-3}
 & total & 217 \\\hline
\multicolumn{2}{c}{TOTAL (Fig.5+Fig.6+anomalous)} & $133\pm 70$ \\\hline\hline
\multicolumn{2}{c}{experimental value \cite{Amsler:2008zz}} & $640\pm246$ \\
\end{tabular}
\end{center}
\caption{Various contributions to the $a_1^+(1260)\rightarrow\pi^+\gamma$ decay width in units of keV.}
\label{tab:width_a1}
\end{table}

\begin{table}[htbp]
\begin{center}
\begin{tabular}{crr} 
\multicolumn{3}{l}{$b_1^+(1235) \rightarrow \pi^+ \gamma$} \\\hline\hline
tree level with model of Refs.~\cite{Roca:2006am,Nagahiro:2008zz} & $\phi$ & 20 \\
 & $\omega$ & 14 \\
 & $\rho$ & - \\\cline{2-3}
 & total & 67 \\\hline\hline
loops type Fig.5 & $K^*K$ & 26 \\\hline
loops type Fig.6 & $K^*K$ & 57 \\\hline
\multicolumn{2}{c}{total (Fig.5+Fig.6)} & 159 \\\hline
loops anomalous & $\omega\pi$ & 2.2 \\
 & $\rho\eta$ & 0.6 \\
 & $\bar{K}^{*0} K^+ (\omega)$ & $6.3\times10^{-2}$ \\
 & $\bar{K}^{*0} K^+ (\phi)$ & 0.1 \\
 & ${K^*}^+ \bar{K}^0 (\omega)$ & $6.3\times10^{-2}$ \\
 & ${K^*}^+ \bar{K}^0 (\phi)$ & 0.1 \\\cline{2-3}
 & total & 4 \\\hline
\multicolumn{2}{c}{TOTAL (Fig.5+Fig.6+anomalous)} & $209\pm 90$ \\\hline\hline
\multicolumn{2}{c}{experimental value \cite{Amsler:2008zz}} & $230\pm60$ \\
\end{tabular}
\end{center}
\caption{Various contributions to the $b_1^+(1235)\rightarrow\pi^+\gamma$ decay width in units of keV.}
\label{tab:width_b1}
\end{table}

\section{Conclusions}

We have developed the formalism to evaluate the radiative decay of axial vector
mesons from the perspective that these states are composite particles of 
pseudoscalar and vector mesons, using the hidden gauge formalism for the
interaction of vector mesons and pseudoscalars among themselves and with 
external sources. 
The formalism is rather rewarding. It shows a clear path to
proceed and
allows the interpretation of the tree level diagrams which
in other formalisms are more difficult to integrate within the corresponding
scheme. Also, one finds finite
radiative decay widths which we compare 
with present experimental results considering theoretical and experimental
uncertainties. We found good results for the radiative decay of the $b_1(1235)$
resonance, while not so good for the $a_1(1260)$ resonance, where large
cancellations occurred in the theoretical framework. Because of these large
cancellations of the main mechanism, extra corrections stemming from the
consideration of loops involving anomalous couplings and extra vector mesons
gave a non-negligible contribution to the radiative decay width.  
The improvement is, however, not remarkable.  
For the case of $b_1$ the agreement of the basic
mechanism with data was already rather good, while the anomalous 
contributions played only a minor role.  


   The finite results obtained within the hidden gauge formalism and the 
simplicity of the approach make the use of this formalism practical and
advisable in this kind of problems. 
Even with the results for the $a_1$ resonance, the agreement with the 
data can be considered fair from the perspective that differences for 
radiative decays between models, and hence with data, are typically of 
the order of one or two orders of magnitude~\cite{Kalashnikova:2005zz}.

  At the same time we have described the formalism with sufficient detail to use
many of the results for related problems, like the interaction of vector mesons
with pseudoscalars, linking the results to existing chiral Lagrangians, the
interaction of vector mesons with themselves, etc. These results should prove
useful in future work dealing with the interaction of vector mesons,
which so far has not received much attention, and were one might expect that
equally interesting results as found in other areas are lying ahead.

\section*{Appendix}

Vertices involving $\rho$, $\pi$ and real photons. 
($\epsilon$, $\epsilon'$, polarization vectors for $\rho$, 
$\epsilon^{(\gamma)}$ for the photon).

\begin{figure*}
     \centering
     \subfigure[]{
          \label{fig:app1}
          \includegraphics[width=.2\linewidth]{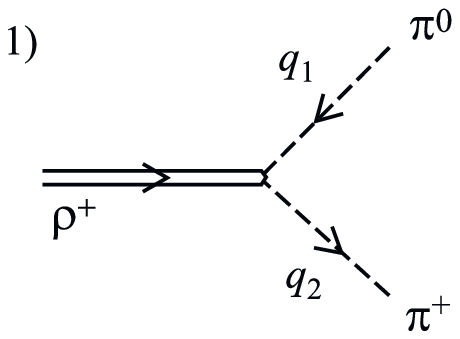}}
     \subfigure[]{
          \label{fig:app2}
          \includegraphics[width=.2\linewidth]{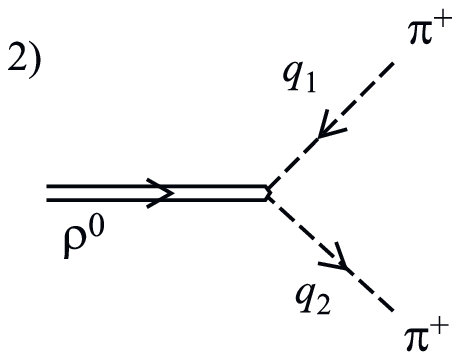}}
     \subfigure[]{
          \label{fig:app3}
          \includegraphics[width=.2\linewidth]{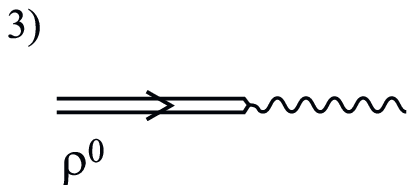}}
\\
    \subfigure[]{
          \label{fig:app4}
          \includegraphics[width=.2\linewidth]{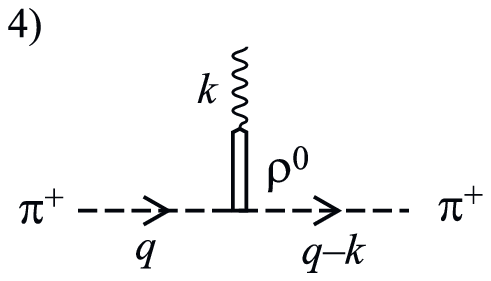}}
     \subfigure[]{
          \label{fig:app5}
          \includegraphics[width=.2\linewidth]{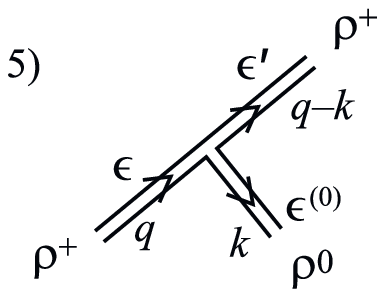}}
     \subfigure[]{
          \label{fig:app6}
          \includegraphics[width=.2\linewidth]{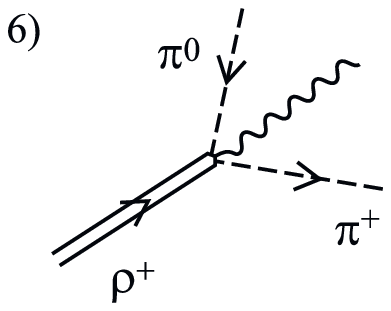}}
\\
    \subfigure[]{
          \label{fig:app7}
          \includegraphics[width=.2\linewidth]{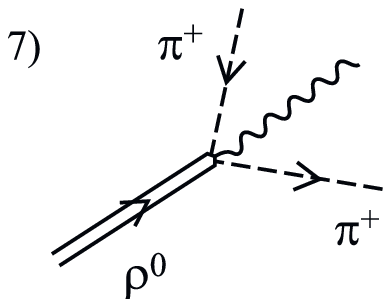}}
     \subfigure[]{
          \label{fig:app8}
          \includegraphics[width=.2\linewidth]{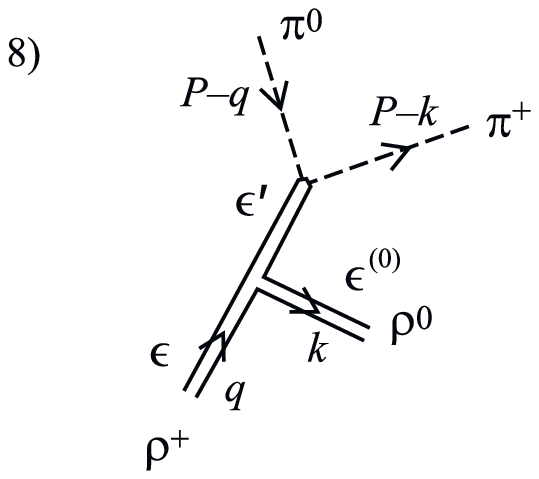}}
     \subfigure[]{
          \label{fig:app9}
          \includegraphics[width=.2\linewidth]{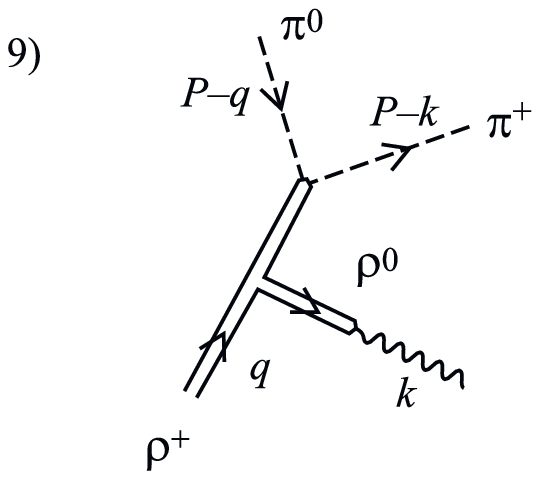}}
     \caption{}
     \label{fig:app}
\end{figure*}

\be
-it^{Fig.\ref{fig:app1}}=-i\frac{M_V^2}{2\sqrt{2}gf^2}(q_1+q_2)
\cdot\epsilon,
\ee

\be
-it^{Fig.\ref{fig:app2}}=i\frac{M_V^2}{2\sqrt{2}gf^2}(q_1+q_2)
\cdot\epsilon,
\ee

\be
-it^{Fig.\ref{fig:app3}}=-ie\frac{M_V^2}{\sqrt{2}g}
\epsilon\cdot\epsilon^{(\gamma)},
\ee

\be
-it^{Fig.\ref{fig:app4}}=ie
(q+q-k)\cdot\epsilon^{(\gamma)},
\ee

\ba
-it^{Fig.\ref{fig:app5}}&=&-i\sqrt{2}g[
(k_\mu\epsilon^{(0)}_\nu -k_\nu \epsilon^{(0)}_\mu)\epsilon^\mu \epsilon'^\nu \nn \\
&+&(-q_\mu \epsilon_\nu +q_\nu \epsilon_\mu)\epsilon'^\mu \epsilon^{(0)\nu}
+((q-k)_\mu\epsilon'_\nu -(q-k)_\nu \epsilon'_\mu)\epsilon^{(0)\mu} \epsilon^\nu],
\ea

\be
-it^{Fig.\ref{fig:app6}}=-ie\frac{M_V^2}{2\sqrt{2}gf^2}\epsilon\cdot\epsilon^{(\gamma)},
\ee

\be
-it^{Fig.\ref{fig:app7}}=ie\frac{\sqrt{2}M_V^2}{2gf^2}\epsilon\cdot\epsilon^{(\gamma)},
\ee

\ba
-it^{Fig.\ref{fig:app8}}&=&-\frac{i}{2f^2}[
(2k-q)\cdot\epsilon (2P-k-q)\cdot\epsilon^{(0)} \nn \\
&-&(k+q)\cdot(2P-k-q) \epsilon\cdot\epsilon^{(0)}
+(2q-k)\cdot\epsilon^{(0)}(2P-k-q)\cdot\epsilon],
\ea
 (neglecting $(q-k)^2/M_V^2$)

\ba
-it^{Fig.\ref{fig:app9}}&=&-e\frac{i}{2\sqrt{2}gf^2}[
(2k-q)\cdot\epsilon (2P-k-q)\cdot\epsilon^{(\gamma)}\nn \\
&-&(k+q)\cdot(2P-k-q) \epsilon\cdot\epsilon^{(\gamma)}
+(2q-k)\cdot\epsilon^{(\gamma)}(2P-k-q)\cdot\epsilon],
\ea
(neglecting $(q-k)^2/M_V^2$).

\section*{Acknowledgments}  

This work is partly supported by DGICYT contract number
FIS2006-03438 and  the JSPS-CSIC
collaboration agreement no. 2005JP0002, and Grant for Scientific
Research of JSPS No.188661.
H.N. is supported by JSPS Research Fellowship for
Young Scientists. 
A. H. is supported
in part by the Grant for Scientific Research Contract
No. 19540297 from the Ministry of Education, Culture,
Science and Technology, Japan.
This research is  part of
the EU Integrated Infrastructure Initiative  Hadron Physics Project
under  contract number RII3-CT-2004-506078.

\end{document}